\documentclass[11pt,twoside]{article}

\usepackage{asp2006}
\usepackage{epsf}
\usepackage{lscape}
\usepackage{graphicx}

\begin{document}
\title{On the fraction of X-ray obscured quasars}   
\author{Junxian Wang, Peng Jiang, Zhenya Zheng, Tinggui Wang}   
\affil{Center for Astrophysics, University of Science and Technology of China, Hefei, Anhui 230026, P. R. China jxw@ustc.edu.cn}    
\author{and the CDF-S team}
\begin{abstract} 
Various studies have claimed that the fraction of obscured AGN drops with 
luminosity, but contrary results are also reported. We present our recent 
studies on the fraction of X-ray obscured quasars in Chandra Deep Fields and 
in the local universe showing that most quasars in CDFs (at redshift of 
1 $\sim$ 4) are obscured and
a consistent pattern in the local universe.
\end{abstract}
\section{Introduction}
A large population of obscured powerful quasars called type-2 quasars has
long been predicted by the widely accepted unified model for active galactic
nuclei. 
They have been predicted to be the
same as
type-1 quasars but with strong obscuration in both optical and soft X-ray band,
due to
the presumable torus.
Most of such type-2 quasars, which might dominate the black hole growth
have been missed by optical surveys
for quasars.
Although doubt has been expressed at times on the existence of
type-2 quasars, mainly due to their
rareness and the fact that they are easily mimicked by other types of AGNs,
hard X-ray surveys have revealed a 
large number of such sources. 
This is mainly because hard X-ray emission is less biased by the obscuration
(but note for Compton-thick absorption, even hard X-ray emission is strongly
attenuated).
Another advantage of X-day data is that X-ray spectral fitting can yield
intrinsic luminosity and absorption simultaneously, which are both
required to study the population of obscured quasars.
Meanwhile infrared surveys also made significant 
progress on discovering such sources (e.g., Mart\'\i nez-Sansigre et al. 2005).

Various studies have shown that the fraction of obscured AGN decreases with
increasing luminosity, suggesting that different from Seyfert galaxies, most 
quasars are unobscured (e.g. Ueda et al. 2003,Steffen et al. 2003; Barger et al.
2005).
For instance, Ueda et al. (2003) computed the X-ray luminosity
function for 2 -- 10 keV selected AGN samples, and found that the fraction of
X-ray absorbed AGNs (with $N_H$ $>$ 10$^{22}$ cm$^{-2}$) drops from $\sim$
0.6 at intrinsic $L_X$ around 10$^{42}$ ergs s$^{-1}$ to around 0.3
at $L_X$ above 10$^{44}$ ergs s$^{-1}$.
However, contrary results are also reported suggesting most quasars are
obscured ones (see Eckart et al. 2006, Dwelly et al. 2005, Dwelly \& Page 2006,
Mart\'\i nez-Sansigre et al. 2005, 2006).
For example, Eckart et al. (2006) found that half of the AGNs identified by the
SEXSI program are X-ray obscured with $N_H$ $>$ 10$^{22}$ cm$^{-2}$,
and the fraction of obscured AGNs is independent of the unobscured luminosity.

\section{Soft gamma-ray selected AGN samples in the local universe}
Soft gamma-ray selected AGN samples (Markwardt et al. 2005; Bassani et al. 
2006) obtained by $SWIFT$ and $INTEGRAL$
provide the best opportunity to study the fraction of obscured AGN in the local
universe in the least biased way.
They have claimed based on these samples, that the
fraction of obscured AGNs in the local universe significantly decrease with
increasing luminosity, and most luminous AGNs are unobscured.

\begin{figure}
\includegraphics[scale=0.75]{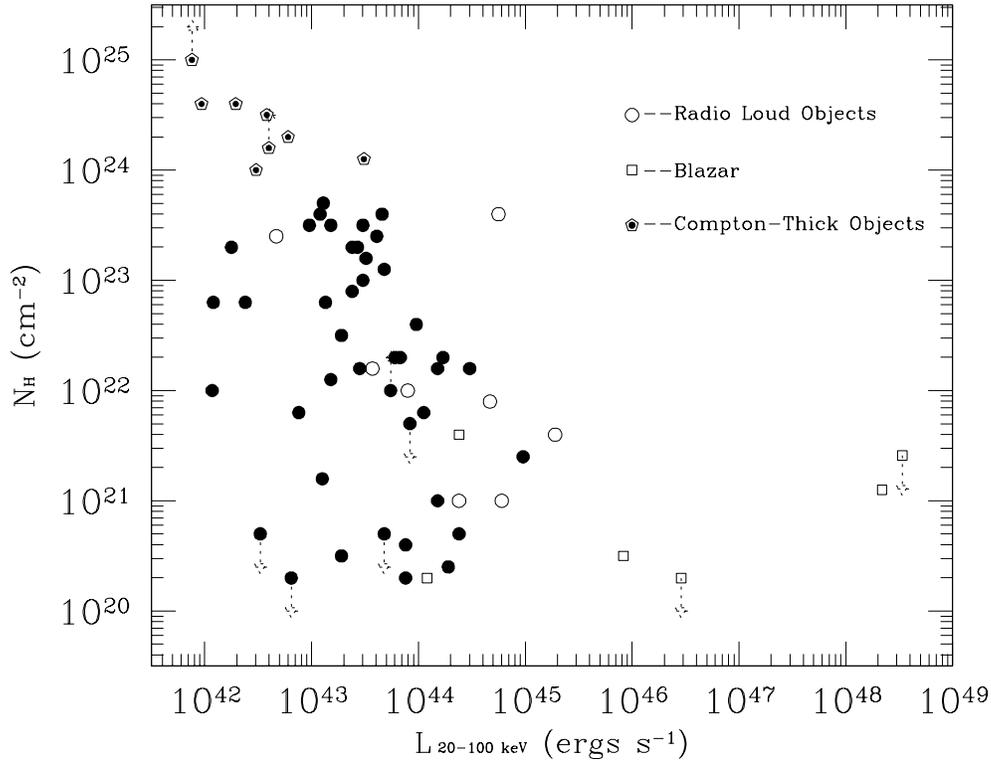}
\caption{
X-ray absorption column density $N_H$ versus soft gamma-ray luminosity
in the 20 -- 100 keV band for $SWIFT$ and INTEGRAL selected AGNs.
See Wang \& Jiang 2006 for details.}
\end{figure}

However, we point out that two correction have to be made to the samples
to study of the fraction of obscured quasars. The
corrections are: a) radio loud AGNs have to be excluded since their X-ray
emission might be dominated by the jet component thus the measured luminosity
and the obscuration does not reflect the intrinsic values in the nuclei;
b) Compton thick sources have to be excluded too since their
soft gamma ray emission are also strongly attenuated by Compton scattering
and their intrinsic luminosities are hard to estimate.

In Fig. 1 we plot the X-ray absorption column density $N_H$ versus soft 
gamma-ray luminosity for a combined sample of AGN selected by $SWIFT$ and 
$INTEGRAL$. We can clearly the clear anti-correlation between the 
X-ray absorption and luminosity is mainly due to the contribution
of radio loud sources and Compton-thick sources, excluding which would 
reduce the significance level of the anti-correlation from $>$ 99.99\%
to 94\%. This indicate that the decrease in the fraction of obscured AGNs
with luminosity in the local universe is not solid.
Larger samples
are required to reach a more robust conclusion.

\begin{figure}
\includegraphics[scale=0.75]{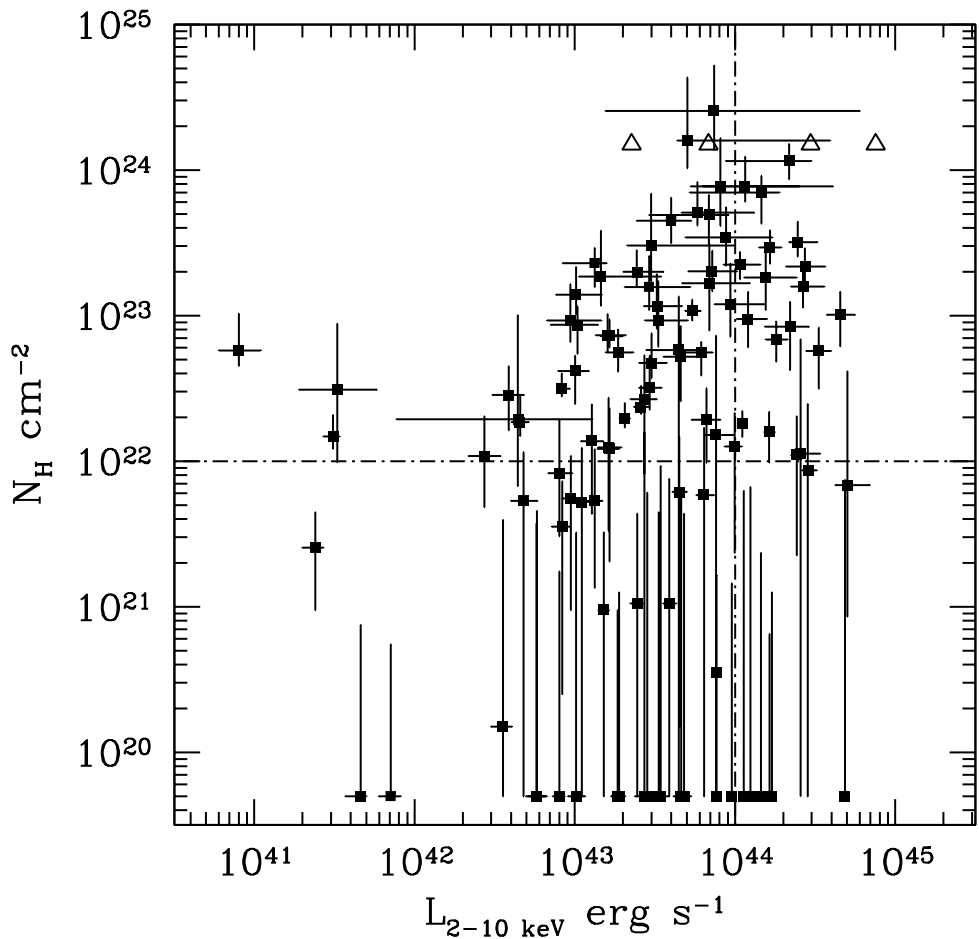}
\caption{
X-ray absorption column density $N_H$ versus absorption-corrected luminosity
for 4 -- 7 keV band selected sources in CDF-S. It's clear that most of the
quasars (with L$_{2-10keV}$ $>$ 10$^{44}$ erg/s) are obscured in X-ray with
N$_H$ $>$ 10$^{22}$ cm$^{-2}$.
}
\end{figure}

\section{X-ray obscured quasars in Chandra Deep Fields}

The 2 Ms $Chandra$ exposure on $Chandra$ Deep Field North (CDF-N,
Alexander et al. 2003) and 1 Ms 
exposure on $Chandra$ Deep Field South (CDF-S, Giacconi et al. 2002) are the deepest X-ray images 
ever taken.
With a redshift completeness of 99\%,
the X-ray sources detected in $Chandra$ Deep Field South (CDF-S)
provide the best sample for the study of the population of obscured quasars.

In a dedicated paper (Wang et al. 2007), we study the population of obscured 
quasars in CDF-S by choosing the 4 -- 7 keV selected sample. We show that
4 -- 7 keV selected sample is even less
biased by the intrinsic X-ray absorption comparing to the normal hard hard 
(2 -- 7 keV). The samples also filter out most of the X-ray faint
sources with too few counts, for which the measurements of
N$_H$ and L$_X$ have very large uncertainties.
Simply adopting the best-fit L$_{2-10keV}$ and N$_H$, we find 71$\pm$19\%
(20 out of 28) of the quasars (with intrinsic L$_{2-10keV}$ $>$ 10$^{44}$
erg s$^{-1}$)
are obscured with N$_H$ $>$ 10$^{22}$ cm$^{-2}$.
Taking account of the uncertainties in the measurements of both N$_H$ and
L$_X$, conservative lower and upper limits of the fraction are
54\% (13 out 24) and 84\% (31 out 37).
In $Chandra$ Deep Field North, the number is 29\%, however, this is mainly due tothe redshift incompleteness.
We estimate a fraction of $\sim$ 50\% - 63\% after correcting
the redshift incompleteness with a straightforward approach. 
Our results
robustly confirm the
existence of a large population of obscured quasars at redshift of 1 $\sim$ 4.

\section{Conclusions}
To study the fraction of obscured AGNs as a function of luminosity is a
difficult issue. It requires large and redshift complete samples. Studies
have tried to combine deep surveys with wide area surveys to cover a
large range of luminosity and redshift. However note most deep surveys suffer
redshift incompleteness in their samples, thus could bring
obvious biases. Using CDF-S where we have a redshift completeness of 99\%,
we find most quasars selected at $z$ = 1 $\sim$ 4 are obscured. 
We also show that in the local universe the previous claimed decrease in the
fraction of obscured AGN with luminosity is mainly due the radio loud sources
and Compton thick sources in the samples, for which the observed luminosities
are not intrinsic. Excluding these sources would erase the decrease
statistically and support a constant fraction of obscured AGNs with luminosity.
This indicates that most quasars at low redshift could possibly also be obscured.




\end{document}